# Freemium Model for Information Provision

Igal Milchtaich, Department of Economics, Bar-Ilan University, Israel
igal.milchtaich@biu.ac.il https://sites.google.com/view/milchtaich

November 2026

**Abstract** The paper explores a theoretical freemium model for the sale of information, drawing on mathematical tools used in the study of repeated zero-sum games and Bayesian persuasion. Unlike standard Bayesian persuasion models, the information seller (IS) is indifferent to the actions taken by the information buyer (IB) and is concerned solely with maximizing the revenue from selling information. Offering some information for free may increase the IB's willingness to pay for additional information.

The information that the IB seeks is about the state of the world. Initially, the IB only knows the prior distribution over possible states. The IS supplies both free and paid information through signals whose state-dependent distributions determine the IB's posterior via Bayes' rule. The IB's utility is a function of the posterior. An optimal free signal is one that maximizes the IS's expected revenue from the subsequent paid signal. That revenue is equal to the IB's expected utility gain when moving from the posterior induced by the free signal to that induced by the paid signal.

The paper characterizes the optimal free and paid signals and derives a formula for the maximal revenue in terms of the IB's utility function. It shows that a revenue gain for the IS from the provision of free information is accompanied by a loss to the IB.

Whether free information can increase the IS's revenue depends on the form of the IB's utility function. In the two-state case, that dependence is fully characterized. In the general case, only necessary conditions are obtained. In particular, if the IB's utility function is convex, the IS can never profit from providing free information. This occurs, in particular, when the IB uses the information to solve a decision problem. By contrast, when the IB is engaged in a strategic interaction with a third party, the IS may benefit from providing free information.

# 1 Introduction

Many products and services, such as software, online newspapers and video games, are marketed using a freemium business model, whereby basic features are offered for free while additional features and capabilities are provided only for a fee. This paper concerns a freemium model for the sale of information. The information provided consists of facts or forecasts grounded in factual knowledge. It may concern such matters as market conditions, public opinion or the weather.

Consider, for example, a weather forecaster that posts forecasts of tomorrow's precipitation probability or sea conditions on a company's website. To get more detailed, say hourly, information, customers need to pay. The logic for doing this may be that the coarse information provided for free incentivizes consumers to pay for additional information. For example, on breezy days, knowing at which hours the sea is expected to be relatively calm may be particularly important. Whether this logic is valid is one question this paper deals with.

The mathematical tools used for studying free provision of information are well known. They were developed for the study of repeated zero-sum games with one-sided private information (Aumann and Maschler 1995; see also Zamir 1992) and later extensively employed in the context of Bayesian persuasion (Kamenica and Gentzkow 2011). The problem here differs from Bayesian persuasion in several important ways. First, the *information seller* (IS) does not try to persuade the *information buyer* (IB) to do anything. What the IB does with the purchased information is of no concern to the IS, who is only interested in maximizing the revenue from selling it. Second, the IB's actions need not even be explicitly specified. The only thing that matters is the utility for the IB from the information, which is a function of the posterior on the possible states of the world that the provided information induces through Bayesian updating. The interaction between the information seller and the information buyer is totally non-strategic.

**Example 1** In the motivating example of Kamenica and Gentzkow (2011), a judge convicts a defendant if and only if there is at least 50% chance that the defendant is guilty. A prosecutor gets payoff 1 from conviction and 0 from acquittal. Suppose that (unlike in the original example) the prior probability $\mu_0$(guilty) that the defendant is guilty is 0.8 and that the prosecutor cannot himself conduct any investigations that can shed light on the defendant's guilt or innocence. However, a third party, the IS, can investigate and may offer to do so for a fee paid by the prosecutor — all with the full knowledge of the judge, with whom the information is shared. The utility $u(\mu)$ that the prosecutor gets from every posterior $\mu$ is shown in Figure 1.

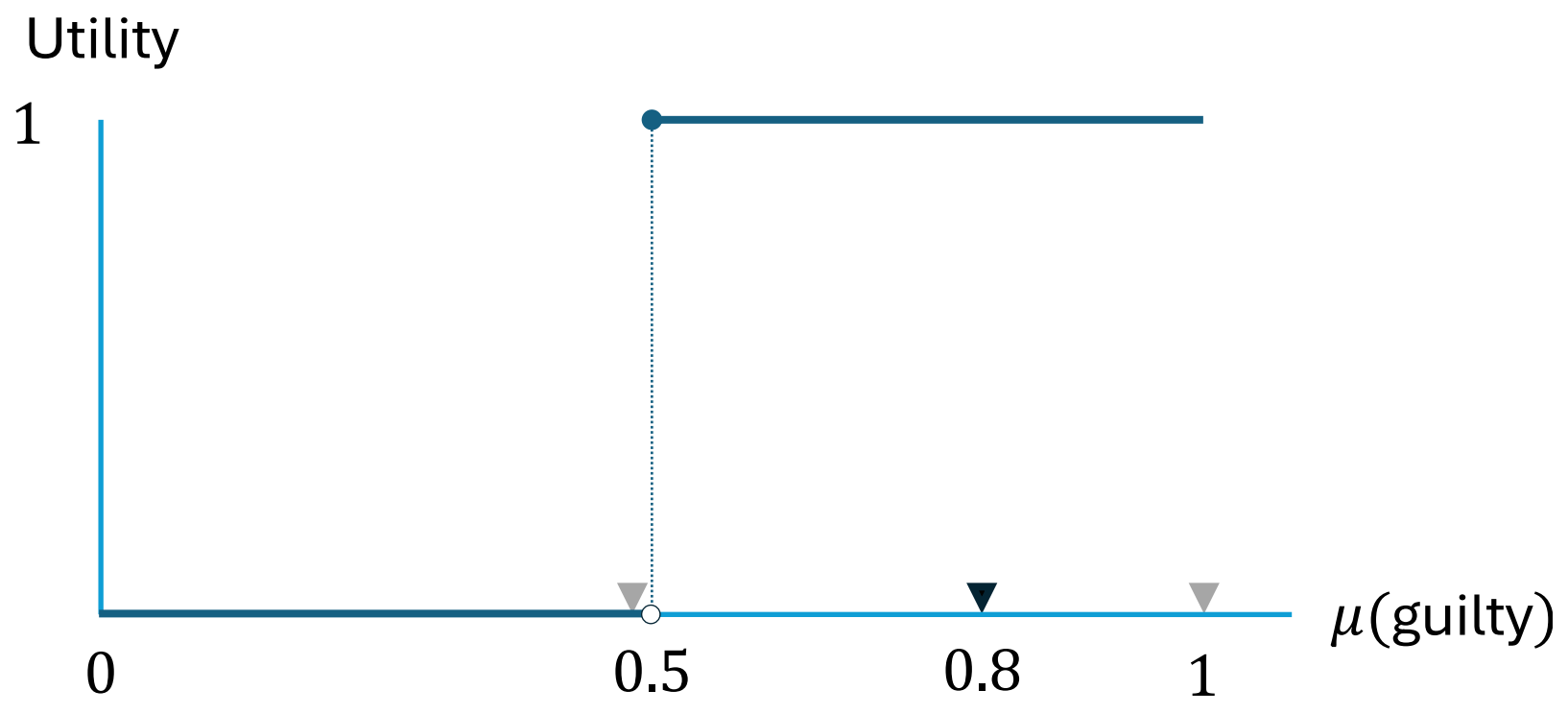

*Figure 1 The prosecutor's utility in Example 1 as a function of the posterior. Shown also are the prior probability $\mu_0$(guilty) $= 0.8$ (black triangle) and the two possible posteriors after the IS's unsolicited announcement (gray triangles).*

Initially, the prosecutor would not be willing to buy any information, as $u(\mu_0) = 1$ is the highest possible utility. However, the IS can incentivize the prosecutor to do so by first providing the following unsolicited information: announcing "Innocent" if the defendant is innocent, and either "Innocent" or "Guilty", the first with probability $0.25 - \epsilon$ and the second with probability $0.75 + \epsilon$, if the defendant is guilty, where $\epsilon$ is some small positive number.[1] If "Guilty" is announced, then the defendant is unequivocally guilty and there is no more information to sell. If "Innocent" is announced, which happens with probability

$$\mu_0(\text{innocent}) + (0.25 - \epsilon\ )\mu_0(\text{guilty}) = (1 - 2\epsilon) \cdot 0.4,$$

then, according to Bayes' rule, the posterior, $\mu_1$, has

$$\frac{\mu_1(\text{guilty})}{\mu_1(\text{innocent})} = \frac{0.8}{0.2} \cdot \frac{0.25 - \epsilon}{1} = 1 - 4\epsilon. \tag{1}$$

As such odds mean acquittal, the prosecutor will be willing, in this case, to pay for additional information that may cast doubt on the defendant's innocence. In particular, the IS may offer the following: announcing "Guilty" if the defendant is guilty, and either "Guilty" or "Innocent", the former with probability $1 - 4\epsilon$ and the latter with probability $4\epsilon$, if the defendant is innocent.[2] "Innocent" unequivocally indicates innocence, while a "Guilty" announcement will lead to conviction because the posterior after such an announcement, $\mu_2$, has

$$\frac{\mu_2(\text{guilty})}{\mu_2(\text{innocent})} = (1 - 4\epsilon) \cdot \frac{1}{1 - 4\epsilon} = 1.$$

The price that the prosecutor will be willing to pay for this information is therefore

---

[1] "Guilty" and "Innocent" are only suggestive labels. For example, the IS may examine fingerprints found at the crime scene and make these announcements if they are or are not, respectively, the defendant's fingerprints.

[2] For example, the IS may offer to try to corroborate the defendant's claimed alibi, with "Innocent" and "Guilty" denoting success or failure, respectively, in doing that.

the probability of a "Guilty" announcement, which, in view of (1), is given by

$$\mu_1(\text{guilty}) + (1 - 4\epsilon)\mu_1(\text{innocent}) = \frac{1 - 4\epsilon}{1 - 2\epsilon}.$$

The expected revenue for the IS from first providing the free information and then the paid information is therefore

$$(1 - 2\epsilon) \cdot 0.4 \cdot \frac{1 - 4\epsilon}{1 - 2\epsilon} = (1 - 4\epsilon) \cdot 0.4.$$

Choosing a sufficiently small $\epsilon$ makes this revenue arbitrarily close to $0.4$.

It is shown below that the IS cannot make more than $0.4$. This amount is the supremum of what can be gained from providing both free and paid information.

The rest of the paper is structured as follows.

Section 2 describes the paper's basic formal framework.

The main differences and similarities between this model and several earlier ones studied in the literature are presented in Section 3.

Section 4 formalizes information provision as signaling schemes, or experiments. It also describes the well-known equivalence between such schemes and "splits" of the prior into several possible posteriors.

The question of the information seller's optimal signaling scheme is tackled in Section 5. The answer gives a simple formula for the maximum possible revenue from the sale of information.

Section 6 examines the possible gain from the provision of free information in the form of an earlier signal. It presents a formula for the maximum possible gain.

Section 7 shows that any gain for the information seller from the provision of free information must come at the expense of the information buyer. The effect on social welfare is therefore never positive.

Whether any gain from the provision of free information is possible depends on the functional form of the information buyer's utility function, which describes the connection between utility and the posterior. Theorem 1 in Section 8 gives both a necessary condition for this and a sufficient condition. In the case of two states of the world, it presents a complete characterization of the utility functions allowing for gain from free information. Theorem 2 in Section 8.1 shows that the latter characterization extends to any utility function that only depends on the expectation of some real-valued function on the set of states.

The necessary condition in Theorem 1 does not hold, and the information seller therefore cannot gain from providing free information, if the utility function of the information buyer is convex. The reason, as Section 8.2 explains, is that a convex utility function means that the IB is never harmed by knowing more rather than less, and in particular, by receiving free information. However, as indicated, gain for the IS from free information must be accompanied by loss to the IB. A convex utility function therefore precludes any possibility of gain.

The utility function is automatically convex if the goal of the information buyer is to solve a decision problem, that is, to identify the optimal action when the outcome of each action is state-dependent. The conclusion, in Section 8.3, is that, if the IB uses the information to solve a decision problem, there can be no gain for the IS from providing information for free.

The last conclusion does not extend to the case where the information buyer is involved in a multi-agent strategic interaction, or game, where the outcome of choosing an action depends also on the other agents' actions. Section 9 gives an example of a two-player investment game with incomplete information where an information seller can gain from providing free information to the players before offering them more information for a fee.

The possibility of gaining from the provision of free information does not depend on the equivalence between signaling schemes and splits of the prior and it may exist even if the information provider can only use deterministic signals. Earlier work (Alkoby et al. 2017, 2019) demonstrated possible gain from free information when the information that the IS sells is the exact state of the world and free information is provided in the form of a set of states, which includes the actual one. Put differently, the IB only learns from the information provided for free that the state is *not* one of several specified possible states. Using an example from Alkoby et al. (2017), Section 10 shows that, in a second-price auction, an information provider limited to deterministic signals may induce the auctioneer to pay more for information about the state of the world (which affects the bidders' valuations) by first providing some information for free.

Several variations, or extensions, of the basic model are considered in Section 11.

In the first variation, studied in Section 11.1, the IS does not release free information before offering paid information but rather uses the threat of releasing damaging information to force the IB to pay more for the information provided for a fee. It is shown that the revenue gain from such extortion is always at least as high as that from the provision of free information, and it can be higher.

The second variation concerns the possibility of repeating the cycle of free–paid information provision. The information seller may find it profitable to repeat this cycle indefinitely. Indeed, the example in Section 11.2 shows that the expected gain from doing so may be unbounded. This possibility may not be very realistic, however, as it requires a completely myopic information buyer.

The third variation, presented in Section 11.3, is an insurance application where the IS would like to sell to an insurance company information about a potential client. Unlike the basic model, here the insurer's expected profit is not determined solely by the posterior. This is because the company does not share the information provided by the IS with the client, whose posterior is therefore different from the company's posterior. It is shown that, because of the client's risk aversion, the information cannot in fact increase the expected profit of the insurer, and so the latter would not actually be willing to purchase it from the IS.

# 2 Model

$\Omega$ is a finite set of possible *states of the world*. A probability distribution on $\Omega$, $\mu \in \Delta(\Omega)$, assigns a probability $\mu(\omega)$ to each state $\omega \in \Omega$. The *prior* $\mu_0$ is a specified, commonly known element of $\Delta(\Omega)$ that gives the probability of each element of $\Omega$ being the true, actual state. Only one party, the *information seller* (*IS*), can investigate and obtain additional information about the state, as detailed below. These investigations do not cost the IS but also have no direct benefit. The only incentive for the IS is to extract as much revenue as possible from selling his services to the *information buyer* (*IB*). The information provided by the IS changes the probability that the IB assigns to each state of the world being the true state. These probabilities, which are computed according to Bayes' rule, constitute the buyer's *posterior*, $\mu \in \Delta(\Omega)$. Without any information, the posterior coincides with the prior $\mu_0$.

For every posterior $\mu$ there is an associated utility $u(\mu)$ for the IB, which is given by a bounded utility function $u: \Delta(\Omega) \longrightarrow \mathbb{R}$. If the posterior changes from $\mu$ to $\nu$, the IB's utility changes by $u(\nu) - u(\mu)$. If this difference is nonnegative, then it is also the price that the IS can ask for the information leading to the change of posterior.[3] Therefore, roughly speaking, the IS would like $\mu$ and $\nu$ to be such that the difference is as large as possible. If the IS can only choose the information that he offers to sell to the IB, then his control is limited to $\nu$ while $\mu = \mu_0$. However, by first providing information for free, the IS can also change $\mu$. The extent to which this possibility can be exploited is examined below.

[3] The assumption that the IS can extract all the surplus from the provision of the information is both natural (he could make a take-it-or-leave-it offer) and convenient. However, it is not crucial for the analysis below and could be replaced by an assumption that the utility gain of the IB is split in some fixed proportions, so that the IS always gets some fraction $0 < \delta \leq 1$ of it. The only modification this requires is to interpret the "revenue" of the IS as the revenue divided by $\delta$.

# 3 Related literature

The model outlined above is broadly related to the literature on markets for information (Bergemann and Bonatti 2019) but differs from most of that literature in one or more key components.

First, the model diverges from the context of imperfectly informed decision makers as studied, for example, by Bergemann et al. (2018). In that context, the information buyer has some private information that the seller does not possess and uses it together with the additional information purchased from the IS for solving a decision problem. The present model does not assume asymmetric information, and it allows for any use of the purchased information, a generality that turns out to greatly affect the results.

The present setup is also different from that of information holdup (Hörner and Skrzypacz 2016), where the information concerns the seller's own competence and the purpose is to have the expert hired by the information receiver.

More closely related to this paper is Mora and Rodriguez (2026). These authors show that it may be possible for an information seller to increase the revenue by first providing some information for free. However, a crucial assumption in their paper is that the information seller and buyer agree to disagree; they have different priors, which are both common knowledge. If the priors coincide, the seller's optimal strategy is just a single sell of information. The present paper shows that the validity of the latter result critically depends on the assumption that the buyer uses the information to solve a decision problem. As Example 1 demonstrates, it does not hold generally.

Another difference from the present paper is that, in the setting of Mora and Rodriguez (2026), the free information provided by the IS always directly benefits the IB. Here, information that the IS chooses to give for free *never* benefits the IB (see Section 7). This is a significant difference, as it means that free-then-paid information is not merely a special case of selling information sequentially. Rather, the free information must be released without the buyer's consent, for example, by making it public.

A potential gain from free information in the case of different priors, and when the information buyer is faced with an operational decision, is also established by Drakopoulos and Makhdoumi (2023). In their model, the buyer's prior is a misspecified estimate of some parameter that affects the utility from the decision. The information seller is aware of the bias but is uncertain about its magnitude. Without misspecification, i.e., when the seller and the buyer priors are the same, the information seller's optimal strategy is just a one-round selling policy: selling the whole data set to the buyer at the beginning of the interaction.

Independently of this work, Lee (2026) recently studied a model very similar to that presented here, although largely in the context of decision problems. His findings include the formulas in Propositions 1 and 2 below and the assertion in Proposition 3, as well as the conclusion that no gain from free information is possible if the buyer's utility function is either concave or convex (Section 8.2).

# 4 Signals

A general way for the information seller to provide information about the state is to commit to a *signaling scheme* (or *information structure* or *experiment*) $\pi$, which is a map from the state space $\Omega$ to the set $\Delta(S)$ of all probability distributions over a finite[4] set $S$ of possible signals. The probability that signal $s$ is sent when the state is $\omega$ is denoted $\pi(s \mid \omega)$. The total probability $\lambda(s)$ of a signal $s \in S$ is therefore given by

$$\lambda(s) = \sum_{\omega \in \Omega} \mu_0(\omega)\pi(s \mid \omega)\,. \tag{2}$$

If this probability is not zero, then, by Bayes' rule, the posterior probability $\mu_s(\omega)$ that the state is $\omega$ when signal $s$ is sent is given by

$$\mu_s(\omega) = \frac{\pi(s \mid \omega)}{\lambda(s)}\mu_0(\omega). \tag{3}$$

It is easy to see that these posteriors have the *martingale property*

$$\sum_{s \in S} \lambda(s)\mu_s = \mu_0. \tag{4}$$

Conversely, given any collection of posteriors $\{\mu_s\}_{s\in S} \subseteq \Delta(\Omega)$ that is indexed by a finite set $S$, if (4) holds for some probability vector $\lambda \in \Delta(S)$, then there exists a signaling scheme $\pi$ satisfying (2) and (3). That scheme is obtained simply by solving (3) for $\pi(s \mid \omega)$ or setting it to zero if $\lambda(s) = 0$. (If $\mu_0(\omega) = 0$, then $\pi(\cdot \mid \omega)$ is irrelevant and can be chosen arbitrarily.)

Note that condition (4) means that the prior $\mu_0$ lies in the convex hull of the posteriors. Therefore, choosing a signaling scheme can equivalently be described as choosing a *split* of the prior: a presentation of $\mu_0$ as a convex combination of a finite collection of posteriors. The coefficients in this presentation are the probabilities with which the corresponding posteriors arise.

[4] The assumption that there are only finitely many possible signals is not crucial and is made only for the sake of simplicity.

# 5 Optimal signaling

An optimal signaling scheme for the IS maximizes the price that the IB would be willing to pay for it. That price is the IB's expected utility gain from receiving the signal. A signaling scheme is therefore optimal if and only if the IB's expected utility from the associated collection of posteriors is maximal, that is, equal to

$$\sup\left\{\textstyle\sum_{s\in S}\lambda(s)u(\mu_s)\;\middle|\;\begin{array}{l}S \text{ is a finite set, } \lambda\in\Delta(\mathrm{S}),\, \{\mu_s\}_{s\in S}\subseteq\Delta(\Omega),\\ \sum_{s\in S}\lambda(s)\mu_s=\mu_0\end{array}\right\}. \qquad (5)$$

Replacing the constant $\mu_0$ in (5) with a variable $\mu\in\Delta(\Omega)$ makes that expression a function of $\mu$. This real-valued function on $\Delta(\Omega)$ is denoted $\operatorname{cav} u$ (Aumann and Maschler 1995, Zamir 1992).[5,6,7] It is the *concave envelope* (or *concavification*) of $u$: the pointwise smallest concave function on $\Delta(\Omega)$ that is everywhere greater than or equal to $u$. Thus, we have:

**Proposition 1** The highest possible revenue for the IS from selling information to the IB is given by

$$(\operatorname{cav} u - u)(\mu_0).$$

Note that the IS cannot gain from selling information in a piecemeal fashion, charging the IB for one signal and then for another. This is because such multiple signals can always be combined into a single signal that gives the same utility gain to the IB, and therefore the same revenue to the IS, as the two signals together.

To see that this "combination invariance" (Frankel and Kamenica 2019) property holds, suppose that the IS employs a signaling scheme $\pi$ but then, after sending a particular signal $s^*\in S$, employs a second signaling scheme $\pi'$, with a set of signals $S'$, probability vector $\lambda'\in\Delta(S')$ and posteriors $\{\mu'_s\}_{s\in S'}\subseteq\Delta(\Omega)$ such that $\sum_{s\in S'}\lambda'(s)\mu'_s=\mu_{s^*}$. The addition of the second signal changes the expected utility of the IB by

$$\lambda(s^*)\left(\sum_{s\in S'}\lambda'(s)u(\mu'_s)-u(\mu_{s^*})\right).$$

[5] It can be shown, using Carathéodory's theorem, that the definition of cav would not be affected by requiring the cardinality of the set $S$ to be no greater than that of $\Omega$. Thus, there would be no loss of generality in assuming that $S$ *is* $\Omega$, i.e., that the IS signals by pointing to a state (which may or may not be the true state).

[6] A corollary of the fact in footnote 5 is that, if the utility function $u$ is upper semicontinuous, then (i) the supremum in the definition of $\operatorname{cav} u$ is actually attained, i.e., sup can be replaced with max, and (ii) $\operatorname{cav} u$ is a continuous function. See Appendix A.

[7] It follows immediately from the definition that, at the extreme points of $\Delta(\Omega)$, $\operatorname{cav} u = u$.

If this expression is positive, then it is also the expected revenue gain for the IS from asking the IB to pay for the second signal whenever $s^*$ is realized. However, the same revenue gain can be obtained by combining the two signaling schemes. The set of signals in the combined scheme is the disjoint union $(S \setminus \{s^*\}) \uplus S'$. Each signal $s \in S \setminus \{s^*\}$ is sent with probability $\lambda(s)$ and induces the posterior $\mu_s$, and each signal $s \in S'$ is sent with probability $\lambda(s^*)\lambda'(s)$ and induces the posterior $\mu'_s$. It is easy to see that the martingale property holds, and therefore these posteriors can indeed be generated by a single signaling scheme.

# 6 Optimal choice of free information

Free information is provided to the IB in the form of a signal that the IS sends before he offers to provide more information for a fee. The signal is the outcome of a second signaling mechanism that the IS chooses for this purpose. The rationale for providing free information is that doing so changes the parameter in the optimal signaling problem discussed above. The prior $\mu_0$ is replaced with the posterior $\mu$ corresponding to the free signal. The problem of finding an optimal free signaling mechanism is therefore mathematically the same as that for the paid signal, except that the IB's utility $u$ is replaced with the IS's revenue $v$, which is given by

$$v = \operatorname{cav} u - u. \tag{6}$$

This observation establishes the following result.

**Proposition 2** The highest possible revenue gain for the IS from the provision of free information is given by

$$(\operatorname{cav} v - v)(\mu_0).$$

The free-information gain increases the revenue of the IS from the subsequent sale of information to

$$\operatorname{cav} v\,(\mu_0) = \operatorname{cav}(\operatorname{cav} u - u)\,(\mu_0).$$

The function cav $v$ is concave and (by the observation in footnote 7) is zero in the set $E \coloneqq \{\delta_\omega\}_{\omega\in\Omega}$ of extreme points of $\Delta(\Omega)$, where there is no information to provide. In the special case $|\Omega| = 2$, this means that, for either state $\omega$, the revenue of the IS is described by a unimodal (first nondecreasing, then nonincreasing) function of the state's prior probability $\mu_0(\omega)$ that is zero at both ends, 0 and 1.

For the IB's utility function $u$ in Figure 1, the maximum revenue that the IS can get from only selling information, $v = \operatorname{cav} u - u$, is given by $v(\mu) = 2\mu(\text{guilty})$ for $\mu(\text{guilty}) < 0.5$ and $v(\mu) = 0$ for $\mu(\text{guilty}) \geq 0.5$. Therefore,

$$\text{cav}\, v\ (\mu) = 2\min\{\mu(\text{guilty}), \mu(\text{innocent})\}.$$

Setting $\mu(\text{guilty}) = \mu_0(\text{guilty}) = 0.8$ gives that the revenue that the IS in Example 1 can obtain by first providing information for free and then selling additional information has a supremum of $0.4$. The corresponding signaling scheme, which brings the IS arbitrarily close to this upper limit, is presented in Section 1.

# 7 Poisoned apple

Free information cannot simultaneously benefit both the information seller and the information buyer. If its provision is beneficial to the IS, then it is necessarily directly harmful to the IB. "Directly" harmful means that the IB's expected utility after the free information is received and before additional information is purchased is lower than the initial utility. Of course, an implicit assumption here is that the IB cannot opt for "strategic ignorance" (Taneva and Wiseman 2021) and refuse to receive the information provided by the IS. In Example 1, such an option arguably does not exist because the IS can present the free information directly to the judge.

**Proposition 3** Free information results in an expected direct loss for the IB that is greater than or equal to the expected revenue gain for the IS.

*Proof.* Consider any signaling scheme for the provision of free information, as described above. The expected direct loss for the IB is given by $u(\mu_0) - \sum_{s\in S} \lambda(s)u(\mu_s)$ and the expected revenue gain for the IS is $\sum_{s\in S} \lambda(s)v(\mu_s) - v(\mu_0)$. Subtracting the latter expression from the former and using (6) gives

$$\text{cav}\, u\ (\mu_0) - \sum_{s\in S} \lambda(s)\ \text{cav}\, u\ (\mu_s)$$

By (4) and the concavity of cav $u$, this difference is nonnegative. ■

In Example 1, provision of free information yields the IS a revenue gain of $(1 - 4\epsilon) \cdot 0.4$. The direct loss to the IB, the prosecutor, is the probability that the free signal is "Innocent", which is $(1 - 2\epsilon) \cdot 0.4$, slightly more than the IS's gain.

Note that Proposition 3 covers also the cases of negative expected direct loss for the IB (so actually a gain) or negative expected revenue gain for the IS (so actually a loss; obviously a nonoptimal choice). From a wider perspective, the proposition reflects the fact that the social-welfare effect of free information is never positive.[8]

[8] This fact, and the argument in the proof of Proposition 4, apply more generally than Proposition 3 and its proof do. In particular, they apply in the case of deterministic signals (Section 10).

**Proposition 4** Provision of free information can only decrease social welfare or leave it unchanged.

*Proof.* As the IS's revenue comes wholly from transfer, social welfare coincides with the IB's expected utility. As discussed in Section 5, the signaling scheme that is chosen by the IS to convey the paid-only information maximizes that utility. The signals conveying the free and paid information can be viewed as a single signaling scheme whose signals are pairs, and so the corresponding expected utility cannot be higher than the maximum. ∎

It follows from the above findings that the IB could only be better off if the provision of free information was infeasible or not allowed.

# 8 When is free information beneficial?

Whether or not the information seller can gain from the provision of free information depends on both the utility function $u$ of the information buyer and the prior $\mu_0$. The following proposition concerns the role of $u$. Recall that $E$ denotes the set of extreme points of $\Delta(\Omega)$, which are the degenerate posteriors.

**Theorem 1** A necessary condition for the existence of *some* prior at which the IS can gain from providing information for free is that the IB's utility function $u$ is *not*

I. concave, or
II. convex in $\Delta(\Omega) \setminus E$.

In the two-state case, $|\Omega| = 2$, this condition is both necessary and sufficient.[9] A sufficient condition for any $\Omega$ is that $u$ has a discontinuity point in the interior of $\Delta(\Omega)$.[10]

*Proof.* By Proposition 2, a prior as above exists if and only if $\operatorname{cav} v - v$ is not identically zero, equivalently, $v$ is not concave.

If $u$ is concave, then $v = \operatorname{cav} u - u$ *is* concave. Indeed, it is identically zero. If $u$ is convex in $\Delta(\Omega) \setminus E$, then $v = \operatorname{cav} u - u$ is concave there. As $v = 0$ in $E$ (see footnote 7), it follows that $v$ is actually concave in the entire domain $\Delta(\Omega)$: for all $\mu, \mu', \mu'' \in \Delta(\Omega)$ and $0 < \lambda < 1$

$$\mu = \lambda\mu' + (1-\lambda)\mu'' \Longrightarrow v(\mu) \geq \lambda v(\mu') + (1-\lambda)v(\mu''). \quad (7)$$

Indeed, if $\mu', \mu'' \in E$, then the implication holds trivially as $v \geq 0$ everywhere. If

[9] With more than two states, the condition is not sufficient. See the example in Appendix B.

[10] The set of priors where free information is beneficial may not include the discontinuity point itself. For example, if $u(\mu^*) = 0$ for some $\mu^* \in \operatorname{int}\Delta(\Omega)$ and $u = 1$ elsewhere, that set consists of all points in $\operatorname{int}\Delta(\Omega)$ *except* $\mu^*$.

only $\mu'$, say, is in $E$, then for all $0 < \epsilon < 1 - \lambda$

$$\mu = \lambda\mu' + (1-\lambda)\mu'' \Rightarrow \mu = \frac{\lambda}{1-\epsilon}\big((1-\epsilon)\mu' + \epsilon\mu''\big) + \left(1 - \frac{\lambda}{1-\epsilon}\right)\mu''$$
$$\Rightarrow v(\mu) \geq \left(1 - \frac{\lambda}{1-\epsilon}\right) v(\mu'')$$

by the concavity of $v$ in $\Delta(\Omega) \setminus E$ and $v \geq 0$. Taking $\epsilon \to 0$ gives (7), as $v(\mu') = 0$. This proves the necessity of the condition in the first part of the theorem.

The sufficiency of the condition in the last part of the theorem follows from the fact that any concave (or convex) function on $\Delta(\Omega)$ is continuous in $\operatorname{int}\Delta(\Omega)$ (Gale et al. 1968). In particular, $u = \operatorname{cav} u - v$ has the latter property if $v$ is concave. Therefore, if $u$ has a discontinuity point in the interior of $\Delta(\Omega)$, then $v$ is not concave.

To finish proving the middle part of the theorem, suppose that $v$ is concave and $\Omega = \{\omega_1, \omega_2\}$. Because of the concavity of $\operatorname{cav} u$, the limit

$$\tilde{u}_i \coloneqq \lim_{\mu \to \delta_{\omega_i}} \operatorname{cav} u\,(\mu)$$

exists for $i = 1,2$ and satisfies

$$\tilde{u}_i \geq \operatorname{cav} u\,(\delta_{\omega_i}),$$

and the linear function $\varphi: \Delta(\Omega) \longrightarrow \mathbb{R}$ defined by

$$\varphi(\mu) = \mu(\omega_1)\tilde{u}_1 + \mu(\omega_2)\tilde{u}_2$$

satisfies

$$\operatorname{cav} u\,(\mu) \geq \varphi(\mu), \qquad \mu \in \Delta(\Omega) \setminus E.$$

Two cases are possible. If the last weak inequality holds as equality for all $\mu \in \Delta(\Omega) \setminus E$, then

$$u(\mu) = \operatorname{cav} u\,(\mu) - v(\mu) = \varphi(\mu) - v(\mu), \qquad \mu \in \Delta(\Omega) \setminus E.$$

This equality and the assumed concavity of $v$ give that the restriction of $u$ to $\Delta(\Omega) \setminus E$ is convex.

The other possible case is when

$$a \coloneqq \sup_{\mu \in \Delta(\Omega)} (\operatorname{cav} u - \varphi)(\mu) = \sup_{\mu \in \Delta(\Omega)} \operatorname{cav}(u - \varphi)\,(\mu) = \sup_{\mu \in \Delta(\Omega)} (u - \varphi)(\mu) \qquad (8)$$

satisfies $a > 0$. The right-most supremum in (8) is attained at some point in $\Delta(\Omega) \setminus E$. This is because (i) $u - \varphi = \operatorname{cav} u - v - \varphi$ is a linear combination of concave functions and is therefore continuous in $\Delta(\Omega) \setminus E$, (ii) $u - \varphi \leq \operatorname{cav} u - \varphi$ everywhere in $\Delta(\Omega)$, and (iii) for $i = 1,2$,

$$(\operatorname{cav} u - \varphi)(\delta_{\omega_i}) = \operatorname{cav} u\,(\delta_{\omega_i}) - \tilde{u}_i \le 0 \quad \text{and} \quad \lim_{\mu \to \delta_{\omega_i}} (\operatorname{cav} u - \varphi)(\mu) = \tilde{u}_i - \tilde{u}_i = 0.$$

Thus, there is some $\mu^* \in \Delta(\Omega) \setminus E$ such that

$$a = (u - \varphi)(\mu^*) \le (\operatorname{cav} u - \varphi)(\mu^*) \le a.$$

It follows from these inequalities that $v(\mu^*) = \operatorname{cav} u\,(\mu^*) - u(\mu^*) = 0$.

Let $\alpha \coloneqq \min\{\mu^*(\omega_1), \mu^*(\omega_2)\} > 0$. For every $\mu \in \Delta(\Omega)$ there is some $\tilde{\mu} \in \Delta(\Omega)$ such that $\alpha\mu + (1-\alpha)\tilde{\mu} = \mu^*$. Since $v \ge 0$, this equality and the assumed concavity of $v$ give

$$0 \le \alpha v(\mu) \le \alpha v(\mu) + (1-\alpha) v(\tilde{\mu}) \le v(\mu^*) = 0,$$

which proves that $v = 0$ identically. The conclusion means that $\operatorname{cav} u = u$, and so $u$ is concave. ■

## 8.1 Utility that depends only on expectation

The characterization (necessary and sufficient condition) in the middle part of Theorem 1 can be generalized. The generalization concerns any utility function that is “unidimensional” in that it can be expressed as depending only on the expectation of some function on $\Omega$. Specifically, suppose that every state $\omega \in \Omega$ is associated with some value $f(\omega)$. With a posterior $\mu$, the expected value is

$$\varphi(\mu) \coloneqq \sum_{\omega \in \Omega} \mu(\omega) f(\omega).$$

The expectation defines a linear function $\varphi: \Delta(\Omega) \longrightarrow \mathbb{R}$. Conversely, any linear function on $\Delta(\Omega)$ can be presented in this form. The image of such a function, $\operatorname{im} \varphi$, is a closed finite interval in the real line.

**Theorem 2** Suppose that the utility function of the IB can be presented as a composite function $u = g \circ \varphi$, where $\varphi: \Delta(\Omega) \longrightarrow \mathbb{R}$ is a linear function and $g: \operatorname{im} \varphi \longrightarrow \mathbb{R}$ is a bounded function. The following conditions are equivalent:

*a.* There is no prior at which the IS can gain from the provision of free information.

*b.* The function $g$ is
  - I. concave, or
  - II. convex in the interior of $\operatorname{im} \varphi$.

*c.* The utility function $u$ is
  - I. concave, or
  - II. convex in the interior of $\Delta(\Omega)$.

*Proof.* Assume, without loss of generality, that $\operatorname{im} \varphi = [0,1]$, and let $\underline{\omega}, \overline{\omega} \in \Omega$ be such that $\varphi(\delta_{\underline{\omega}}) = 0$ and $\varphi(\delta_{\overline{\omega}}) = 1$. With $\widehat{\Delta} \coloneqq \Delta(\{\underline{\omega}, \overline{\omega}\})$ viewed as a subset of $\Delta(\Omega)$, consider the following restrictions to $\widehat{\Delta}$:

$$\hat{u} := u|_{\hat{\Delta}}, \quad \hat{v} := v|_{\hat{\Delta}}.$$

It is easy to see that (because a split of an element on a face remains on that face) $\operatorname{cav}\hat{u} = (\operatorname{cav} u)|_{\hat{\Delta}}$ and therefore $\hat{v} = \operatorname{cav}\hat{u} - \hat{u}$.

Recall that, by Proposition 2, condition *a* holds if and only if $v$ is concave.

(*a* ⟹ *b*) Suppose that $v$ is concave. Then, $\hat{v}$ is also concave, and so it follows from the middle part of Theorem 1 that

I. $\hat{u}$ is concave, or
II. $\hat{u}$ is convex in the interior of $\hat{\Delta}$.

For every $t', t'' \in [0,1]$ and $0 \leq \lambda \leq 1$, the linearity of $\varphi$ gives that

$$\begin{aligned}
&\lambda g(t') + (1-\lambda)g(t'') - g(\lambda t' + (1-\lambda)t'') \\
&\quad = \lambda g\left(t'\varphi(\delta_{\bar{\omega}}) + (1-t')\varphi(\delta_{\underline{\omega}})\right) + (1-\lambda) g\left(t''\varphi(\delta_{\bar{\omega}}) + (1-t'')\varphi(\delta_{\underline{\omega}})\right) \\
&\quad - g\left(\lambda\left(t'\varphi(\delta_{\bar{\omega}}) + (1-t')\varphi(\delta_{\underline{\omega}})\right) + (1-\lambda)\left(t''\varphi(\delta_{\bar{\omega}}) + (1-t'')\varphi(\delta_{\underline{\omega}})\right)\right) \\
&\quad = \lambda \hat{u}\left(t'\delta_{\bar{\omega}} + (1-t')\delta_{\underline{\omega}}\right) + (1-\lambda)\hat{u}\left(t''\delta_{\bar{\omega}} + (1-t'')\delta_{\underline{\omega}}\right) \\
&\quad - \hat{u}\left(\lambda\left(t'\delta_{\bar{\omega}} + (1-t')\delta_{\underline{\omega}}\right) + (1-\lambda)\left(t''\delta_{\bar{\omega}} + (1-t'')\delta_{\underline{\omega}}\right)\right).
\end{aligned}$$

Therefore, condition I above implies that $g$ is concave, and condition II implies that $g$ is convex in $(0,1)$. Thus, condition *b* holds.

(*c* ⟹ *b*) If $u$ is concave, then so is $\hat{u}$, which as shown above implies that $g$ is concave too. Suppose now that $u$ is convex in the interior of $\Delta(\Omega)$. For every $t', t'' \in (0,1)$ there exist $\mu', \mu'' \in \operatorname{int}\Delta(\Omega)$ such that $\varphi(\mu') = t'$ and $\varphi(\mu'') = t''$. For every $0 \leq \lambda \leq 1$, the assumption on $u$ gives that

$$\begin{aligned}
&\lambda g(t') + (1-\lambda)g(t'') - g(\lambda t' + (1-\lambda)t'') \\
&\qquad = \lambda u(\mu') + (1-\lambda)u(\mu'') - u(\lambda\mu' + (1-\lambda)\mu'') \geq 0.
\end{aligned}$$

The inequality proves that $g$ is convex in the interior of $\operatorname{im}\varphi$.

(*b* ⟹ *c*,*a*) If $g$ is concave, then so are $u = g \circ \varphi$ and $v = \operatorname{cav} u - u \; (= 0)$. If $g$ is convex in $(0,1)$, then $u$ is convex in $\varphi^{-1}\big((0,1)\big)$ (and therefore also in the interior of $\Delta(\Omega)$, which is a subset) and $v = \operatorname{cav} u - u$ is concave there. By the argument used in the proof of Theorem 1 (to establish (7)), to prove that $v$ is concave in its entire domain $\Delta(\Omega)$ it suffices to show that $v = 0$ in $\varphi^{-1}(\{0,1\})$, that is, $\operatorname{cav} u = u$ there. Indeed, whenever any $\mu \in \Delta(\Omega)$ such that $\varphi(\mu) = 0$ or $= 1$ is split into several posteriors, all of them must satisfy a similar equality, which implies that the value of $u$ at each of these posteriors is the same as at $\mu$. It follows that $\operatorname{cav} u\,(\mu) = u(\mu)$. ■

Appendix B shows that the unidimensionality assumption in the theorem cannot be dispensed with. For general utility functions, which cannot be presented as above, condition *a* does not imply *c* or vice versa.

## 8.2 Concave or convex utility

By Theorem 1, if the information buyer has a concave or convex utility function, then the information seller can never gain from providing information for free. This result can be explained by the fact that concavity of $u$ is equivalent to the condition that the IB is never better off having more information rather than less, in the sense of Blackwell order, and convexity is equivalent to the condition that the IB is never worse off.

Indeed, an IB with a concave utility function, who cannot gain from having more information, will never buy it, a fact that cannot be changed by the provision of free information. With a convex utility function, free information cannot harm the IB; the expected change in $u$ is always positive or zero. However, it is shown in Section 7 that, for any $u$, gain to the IS from the provision of free information is necessarily accompanied by harm to the IB. Therefore, if $u$ is convex, no such gain is possible.

## 8.3 Decision problems

An important case in which the utility function $u$ is necessarily convex is when the IB uses the information provided by the IS to solve a *decision problem*. Such a problem involves a set of actions $A$, from which the IB has to choose one. The payoff depends on the chosen action $a$ and on the state of the world $\omega$ and is given by a function $U: A \times \Omega \longrightarrow \mathbb{R}$. With a posterior $\mu$, the expected payoff from choosing action $a$ is $\sum_{\omega\in\Omega} \mu(\omega)U(a,\omega)$. The IB's utility function, $u: \Delta(\Omega) \longrightarrow \mathbb{R}$, reflects a choice of a payoff-maximizing action:

$$u(\mu) = \max_{a\in A} \sum_{\omega\in\Omega} \mu(\omega)U(a,\omega).$$

This function is convex as it is the pointwise maximum of linear functions. Thus, we have:

**Corollary** If the IB uses the information for solving a decision problem, then the IS can never gain from providing information for free.

# 9 Strategic interactions

Free information may be beneficial to the IS if, instead of a single-agent decision problem, the IB is involved in a multi-agent strategic interaction. A strategic interaction, or game, involves several decision makers whose optimal choices of action depend on the others' actions. It is well known that, in this setting, the value of (even private) information can be negative. As explained in Section 8.2, this may give room for the IS to gain from providing information for free.

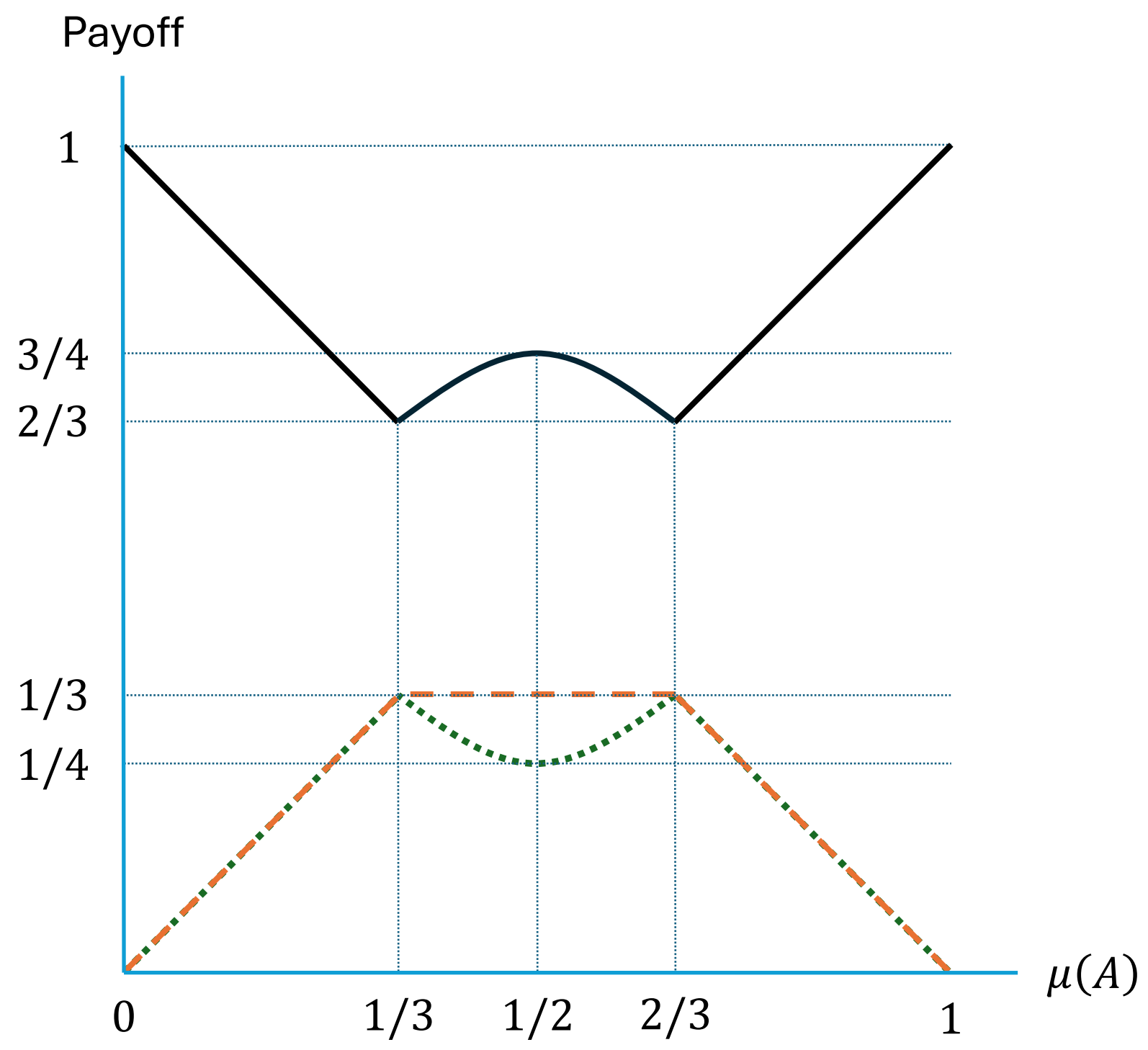

*Figure 2 The symmetric equilibrium payoff $u$ in the investment game in Example 2 as a function of the posterior (solid black line). Also shown are the maximum revenue $v = cav\, u - u$ of the information seller from selling information (dotted green line) and the maximum revenue $cav\, v$ from first providing free information and then paid information (dashed orange line).*

**Example 2** *Investment game.* Two investors play a symmetric $2 \times 2$ game with incomplete information. Each of them chooses whether to invest in firm $A$ or in firm $B$. Only one firm will be successful. The total profit from investment in a firm, which is equally divided between the two investors if they both chose it, is $1$ if the firm is successful and $0$ otherwise. The prior $\mu_0$ is that the two firms are equally likely to be successful, $\mu_0(A) = \mu_0(B) = 1/2$.

An information seller can provide information to the investors about the identity of the successful firm.[11] The information gives rise to a posterior $\mu$. The investors then simultaneously choose where to invest. Their expected profits are given by the payoff matrix (for the row player)

$$\begin{array}{cc} & \begin{array}{cc} A & B \end{array} \\ \begin{array}{c} A \\ B \end{array} & \begin{pmatrix} \mu(A)/2 & \mu(A) \\ \mu(B) & \mu(B)/2 \end{pmatrix}. \end{array}$$

[11] The party to the deal is both investors together. However, it would make little difference if only one investor was offered to buy the information, with the proviso that it cannot be kept private but must be shared with the other investor.

The symmetric $2 \times 2$ game with this payoff matrix has a unique symmetric equilibrium strategy. A firm whose posterior probability of being successful is $\alpha \geq 1/2$ is selected for investment with probability $\min\{1,3\alpha - 1\}$ and the other firm is selected with the complementary probability $\max\{0,2 - 3\alpha\}$. The corresponding equilibrium payoff is $(1/2) \max\{\alpha, 3\alpha(1 - \alpha)\}$. Expressing this symmetric equilibrium payoff as a function of the posterior $\mu$ and multiplying by 2 gives the investors' utility function $u$, which is shown in Figure 2. At the prior, the utility is $u(\mu_0) = 3/4$.

Clearly, $\text{cav}\, u = 1$ identically, and so $v = 1 - u$. By Proposition 1, the IS can get at most $v(\mu_0) = 1/4$ by selling information about the state to the investors. This maximum is achieved by selling full information, that is, revealing the identity of the successful firm, which increases the players' utility to 1. However, by Proposition 2, the IS can achieve a higher revenue of $\text{cav}\, v\ (\mu_0)$ by first providing information for free. It is easy to see that $\text{cav}\, v\ (\mu) = \min\{\mu(A), \mu(B), 1/3\}$, and so this higher revenue is $1/3$. It is obtained by sending a signal that puts the posterior at either $\mu(A) = 2/3$ or $\mu(B) = 2/3$, with equal probability. (The IS can produce such a signal by first finding out the identity of the successful firm and then rolling a die, with four outcomes mapped to pointing at that firm and two to pointing at the other firm.) In both cases, the posterior $\mu$ is such that $u(\mu) = 2/3$, which means that the investors will now be willing to pay $1/3$ to know the state.

# 10 Information provision without randomization

The definition of signaling scheme allows for randomization. A state generally does not determine a signal but a distribution over signals. This setting is crucial for guaranteeing full generality of the associated systems of posteriors. However, randomization is not a crucial element of the argument for free information provision, which may be worthwhile even if only deterministic signals are feasible.

Call a signaling scheme $\pi$ *deterministic* if $\pi(s \mid \omega) \in \{0,1\}$ for all $s$ and $\omega$, in other words, if in each state only one signal can be sent. Such a scheme induces a partition of $\Omega$. Two states are in the same partition element if the signal is the same in both. Thus, what the IB learns from a deterministic signaling scheme is that the true state lies in a particular subset of $\Omega$. The corresponding posterior is obtained by assigning zero probability to states outside the subset and normalizing.

An example of a deterministic signaling scheme is the weather forecast example mentioned in the introduction. A state in this case is a complete forecast, specifying the hourly temperature, say. The daily average maps states into a one-dimensional scale. It is a piece of information that can be provided without completely revealing the state.

The following auction provides another example.

**Example 3** (Alkoby et al. 2017) *Second-price auction*. There are four bidders. Each of them is of type $t_1, t_2$ or $t_3$ and the types are independent. A bidder's valuation of the item being auctioned depends on his type, which the bidder knows, and on the state of the world, which he does not know. There are five states, $\omega_1$ through $\omega_5$. The following table gives the prior $\mu_0$ (second row), the probability of each type (second column) and the bidder's valuation for each type-state pair.

| | | $\omega_1$ | $\omega_2$ | $\omega_3$ | $\omega_4$ | $\omega_5$ |
|---|---|---|---|---|---|---|
| | | 0.28 | 0.19 | 0.20 | 0.07 | 0.26 |
| $t_1$ | 0.38 | 66 | 5 | 35 | 45 | 24 |
| $t_2$ | 0.22 | 72 | 86 | 28 | 73 | 14 |
| $t_3$ | 0.40 | 84 | 14 | 59 | 37 | 81 |

The IB is the auctioneer. However, any information provided by the IS must be shared with the bidders and will thus affect their bids. The utility of the auctioneer is the expected revenue from the auction.

The auction is second-price, and so it is a dominant strategy for the bidders to bid their expected valuations. The expectation is with respect to the posterior, which reflects the information provided by the IS. The auctioneer's revenue is the second-highest bid. Thus, with a posterior $\mu$, the utility $u(\mu)$ is the expected second highest among the four bidders' valuations. For the prior $\mu_0$, computation gives that $u(\mu_0) = 54.1$.

The IS is restricted to using deterministic signaling schemes. Thus, he can only inform the auctioneer (and the bidders) that one or more states are not the true state or, as a special case, point to the true state. Exhaustive search gives that the most profitable signaling scheme is telling whether or not the state is $\omega_3$. This information increases the auctioneer's utility to 54.5. The IS can therefore ask $(54.5 - 54.1 =)$ 0.4 for it.

Suppose, however, that the IS first tells the auctioneer whether or not the state is $\omega_2$. If it is not $\omega_2$, an event that has probability 0.81, then the posterior $\mu$ is such that $u(\mu) = 58.6$. The auctioneer will then be willing to pay to learn the true state. This is because $u(\delta_{\omega_1}) = 77.4$, $u(\delta_{\omega_3}) = 47.3$, $u(\delta_{\omega_4}) = 49.5$ and $u(\delta_{\omega_5}) = 53.6$ (where $\delta_\omega$ is the posterior assigning probability 1 to state $\omega$), and so knowing the state increases the auctioneer's expected utility to $((0.28 \cdot 77.4 + 0.2 \cdot 47.3 + 0.07 \cdot 49.5 + 0.26 \cdot 53.6)/0.81 =)$ 59.9. Providing the free information therefore increases the expected revenue of the IS from $0.4$ to $(0.81 \cdot (59.9 - 58.6) =)$ just over $1$.

Note that in a *first-price* auction in which the bidders' equilibrium strategy is to bid some fixed fraction of their valuation, provision of free information cannot benefit the IS. This is because the auctioneer's profit in this case is proportional to the expected *highest* valuation, which is a convex function of the posterior (see Section 8.2).

# 11 Variations

## 11.1 Extortion

As any free information that increases the revenue of the IS is detrimental to the IB (see Section 7), it seem plausible that the IS might gain also from not releasing free information but only threatening to do so, unless the IB pays him. In other words, the IS may want to use paid information as a carrot and free information as a stick.[12]

The maximum revenue that the IS can extract by extorting the IB in this way is the difference between the highest expected utility for the IB with any signaling scheme and the lowest such utility. The former is given by (5) and the latter is given by a similar expression in which sup is replaced with inf.[13] The difference between the two can be written as

$$(\operatorname{cav} u + \operatorname{cav}(-u))(\mu_0). \qquad (11)$$

As shown in Section 6, the maximum revenue with free-then-paid information is $\operatorname{cav} v\,(\mu_0)$. The next proposition shows that the latter is never higher than (11) , and it can be lower.

**Proposition 5** For every utility function $u$,

$$\operatorname{cav} u + \operatorname{cav}(-u) \geq \operatorname{cav} v\,, \qquad (12)$$

while the reverse inequality does not always hold.

*Proof.* As $v = \operatorname{cav} u + (-u)$, it is (pointwise) less than or equal to the function on the left-hand side of (12). As the latter function is concave, the inequality extends to the concave envelope of $v$.

Consider now any strictly concave $u$ that is zero in $E$, the set of extreme points of $\Delta(\Omega)$. The left-hand side of (12) is equal to $u$, and the right-hand side is

[12] I am indebted to Joel Watson for raising the question of the equivalence between this setting and the one considered in the previous sections.

[13] The latter expression corresponds to the *convex envelope* of $u$.

identically zero. The inequality is therefore strict in all but a finite number of points.[14] ■

With paid information only (Section 5), the expected revenue for the IS is $(\text{cav}\,u - u)(\mu_0)$. Comparison with (11) shows that the gain from extortion is given by the function $\text{cav}(-u) + u$. This function is identically zero if and only if $-u$ is concave, equivalently, $u$ is convex. As shown in Section 8, a convex utility function is *one* case in which the IS cannot gain from the provision of free information regardless of the prior. Here, it is the *only* case. For any nonconvex utility function, there is a prior at which the IS can use extortion to improve on just selling information.

## 11.2 Money pump

The IB's willingness to pay for the information provided by the IS is predicated on the assumption that this information and the posterior it induces are final. If such finality is not guaranteed, the IS may be tempted to repeat the procedure: provide unsolicited information that will make the IB willing to pay for more information. In fact, as the following example shows, if there is no limit on the number of times the IS can repeat this cycle, he may be able to extract an unbounded total payment from the IB.

**Example** 1 (cont.) With the free and paid signaling described above, there is probability $(1 - 4\epsilon) \cdot 0.4$ of ending with the posterior $\mu_2$, for which $\mu_2(\text{guilty}) = \mu_2(\text{innocent}) = 0.5$. Suppose that, in this case, the IS makes an unsolicited announcement of "Innocent" if the defendant is innocent, and "Innocent" with probability $1 - \epsilon_1$ and "Guilty" with probability $\epsilon_1$ if the defendant is guilty, where $0 < \epsilon_1 < 1/2$.

If the announcement is "Innocent", which happens with probability $1 - 0.5\epsilon_1$, the posterior probability that the defendant is guilty is $(1 - \epsilon_1)/(2 - \epsilon_1) < 0.5$, which entails acquittal. The IS can then sell the prosecutor the "opposite" signaling scheme: "Guilty" if the defendant is guilty, and "Guilty" with probability $1 - \epsilon_1$ and "Innocent" with probability $\epsilon_1$ if the defendant is innocent. A "Guilty" announcement restores the fifty-fifty posterior and thus leads to conviction. The price that the prosecutor is willing to pay for the information is therefore equal to the probability of that announcement, which is $(1 - \epsilon_1)/(1 - 0.5\epsilon_1)$.

With probability $1 - \epsilon_1$, the posterior after the second announcement is again $\mu_2$. The IS can then repeat the two-stage cycle, replacing $\epsilon_1$ with increasingly smaller

[14] In this example, the IS cannot sell any information to the IB but can extort the whole IB's utility from the prior by threatening to reveal the state (which would reduce the utility to zero). By contrast, for the utility functions in Examples 1 and 2 the two sides of (12) are equal, meaning that there is no prior at which extortion is better for the IS than free information.

$\epsilon_2, \epsilon_3, \ldots$. For every $n$, the probability that at least $n$ cycles will be completed is $(1-\epsilon_1)(1-\epsilon_2)\cdots(1-\epsilon_n)$, which is greater than $\exp(-2\sum_i \epsilon_i)$. By choosing the epsilons in such a way that the series in the exponent converges, the IS can guarantee that, with positive probability (which can be made arbitrarily close to 1), the posterior will repeatedly return to fifty-fifty, and so the cycle can be indefinitely repeated. As the revenue gain in each iteration is close to 1, this means that the expected revenue is unbounded.

## 11.3 Insurance

The model studied in the previous sections assumes that the utility for the information buyer is completely determined by the posterior. However, this may not be so when a third party is involved and that party, unlike the judge in Example 1 and the bidders in Example 3, does not share the IB's information but only observes the latter's response to it. This lack of a common posterior may complicate the relation between the signaling scheme employed by the IS and the revenue from selling information.

In the following example, the IB is an insurance company that receives from the IS information about an insured person, the third party.[15]

**Example 4** A risk averse person incurs a loss of 1 with probability $0 < \mu_0 < 1$. The person's (Von Neumann-Morgenstern) disutility from any loss or payment $t$ is given by $\bar{u}(t)$, where $\bar{u}$ is a strictly increasing and strictly convex (as it describes *dis*utility) function that is normalized in such a way that $\bar{u}(0) = 0$ and $\bar{u}(1) = 1$.

A risk neutral insurer offers to insure the person against the loss. The highest possible premium the insurer can ask is the solution $p_0$ of the insured person's incentive compatibility equation

$$\bar{u}(p_0) = \mu_0. \tag{13}$$

The corresponding expected profit for the insurer is

$$p_0 - \mu_0.$$

An IS can investigate and shed light on the insured person's likelihood of incurring the loss. The insurer can use the investigation findings to decide whether to offer coverage with some predetermined premium $0 < p < 1$. Coverage will be offered if the posterior $\mu$ set by the findings is such that $\mu < p$ and not offered if $\mu > p$. (If $\mu = p$, the policy may be offered, not offered, or offered with some probability between 0 and 1.) If the policy is acceptable to the person, the insurer's expected profit is

[15] The idea to look at this scenario was suggested to me by Joel Sobel.

$$\mathbb{E}_\mu[[p-\mu]^+].$$

The insured person is aware of the IS's investigation but does not know the findings. However, the very fact that the insurance policy was offered sets the person's posterior to

$$\mathbb{E}_\mu[\mu \mid \text{offer made}].$$

Therefore, the policy is acceptable if and only if

$$\bar{u}(p) \le \mathbb{E}_\mu[\mu \mid \text{offer made}]. \quad (14)$$

By the above description of the insurer's strategy and the martingale condition,

$$\mathbb{E}_\mu[\mu \mid \text{offer made}] \le \mathbb{E}_\mu[\mu] = \mu_0.$$

It therefore follows from Eq. (13) that no premium $p$ with $p > p_0$ (for which $\bar{u}(p) > \bar{u}(p_0)$) can be acceptable to the person. The rest of this analysis concerns the case $p \le p_0$ (hence, $\bar{u}(p) \le \mu_0$).

A signaling scheme is revenue maximizing for the IS if and only if it maximizes the insurer's expected profit subject to the acceptance constraint (14). One such maximizer is the following signaling scheme: a signal inducing the posterior $\mu = \bar{u}(p)$ ($< p$, by the strict convexity of $\bar{u}$ and the normalization at the endpoints) is sent with probability

$$\frac{1-\mu_0}{1-\bar{u}(p)}$$

and a signal giving the posterior 1 ($> p$) is sent with the complementary probability (which may be 0). In the first case, the insurance offer is made and is accepted, and in the second case (where the posterior of 1 means that the loss will happen for sure) insurance is not offered. (Thus, the insured person is effectively informed about the IS's findings.) The expected profit of the insurer is

$$\mathbb{E}_\mu[[p-\mu]^+] = \frac{1-\mu_0}{1-\bar{u}(p)}(p-\bar{u}(p)) = (1-\mu_0)\left(1-\frac{1-p}{1-\bar{u}(p)}\right). \quad (15)$$

To see that this profit is indeed the highest possible, note, first, that under any signaling scheme

$$\Pr(\text{offer made})\,\mathbb{E}_\mu[1-\mu \mid \text{offer made}] \le \mathbb{E}_\mu[1-\mu] = 1-\mu_0.$$

Therefore, for a signaling scheme satisfying the acceptance constraint (14)

$$\begin{aligned}\text{Insurer's expected profit} &= \Pr(\text{offer made})\,\mathbb{E}_\mu[p-\mu \mid \text{offer made}] \\ &\le \frac{1-\mu_0}{\mathbb{E}_\mu[1-\mu \mid \text{offer made}]}\mathbb{E}_\mu[p-\mu \mid \text{offer made}]\end{aligned}$$

$$= (1-\mu_0)\left(1 - \frac{1-p}{1-\mathbb{E}_\mu[\mu \mid \text{offer made}]}\right).$$

By (14), the last expression is less than or equal to (15).

Consider now the dependence of the insurer's maximal expected profit (15) on the premium $p$. The quotient $(1-p)/(1-\bar{u}(p))$ in (15) is the inverse of the average slope of $\bar{u}$ in the interval $[p, 1]$. Because $\bar{u}$ is a strictly convex function, the average slope is a strictly increasing function of $p$, and therefore the same is true for the profit. Thus, the profit with any premium $p < p_0$ is lower than with $p = p_0$. For the latter, the profit is $p_0 - \mu_0$, the profit without the IS's involvement. This finding means that the insurer will not be willing to buy any information from the IS.

Note that, unlike in the previous examples, the IS cannot incentivize the IB to buy information by making some information publicly available, thus effectively changing the prior. This is because the above conclusion, that no information will be bought, holds for any prior.

# Appendix A Continuous concave envelope

**Proposition A1** If the utility function $u$ is upper semicontinuous, then $\operatorname{cav} u$ is a continuous function and it can be presented as

$$\operatorname{cav} u\,(\mu) = \max\left\{ \textstyle\sum_{\omega\in\Omega} \lambda(\omega) u(\mu_\omega) \;\middle|\; \begin{matrix} \lambda \in \Delta(\Omega), \{\mu_\omega\}_{\omega\in\Omega} \subseteq \Delta(\Omega), \\ \sum_{\omega\in\Omega} \lambda(\omega)\mu_\omega = \mu \end{matrix} \right\}. \qquad (16)$$

*Proof.* As remarked in footnote 5, it follows from Carathéodory's theorem that $\operatorname{cav} u$ can always be put in a form similar to (16) except that max is replaced with sup. Therefore, for every sequence $(\mu^n)_{n\geq 1}$ in $\Delta(\Omega)$ converging to a limit $\mu$, there are elements $\lambda^n \in \Delta(\Omega)$ and $\{\mu_\omega^n\}_{\omega\in\Omega} \subseteq \Delta(\Omega)$ for every $n \geq 1$ such that $\sum_{\omega\in\Omega} \lambda^n(\omega)\mu_\omega^n = \mu^n$ and

$$\operatorname{cav} u\,(\mu^n) \leq \sum_{\omega\in\Omega} \lambda^n(\omega) u(\mu_\omega^n) + \frac{1}{n}.$$

The compactness of $\Delta(\Omega)$ guarantees that (by moving to subsequences, if necessary) convergence can be assumed: there exist $\lambda$ and $\{\mu_\omega\}_{\omega\in\Omega}$ such that $\lambda^n \to \lambda$ and $\mu_\omega^n \to \mu_\omega$, $\omega \in \Omega$. Necessarily, $\sum_{\omega\in\Omega} \lambda(\omega)\mu_\omega = \mu$, and if $u$ is upper semicontinuous, then also

$$\limsup_{n\to\infty} \operatorname{cav} u\,(\mu^n) \leq \limsup_{n\to\infty} \sum_{\omega\in\Omega} \lambda^n(\omega) u(\mu_\omega^n) \leq \sum_{\omega\in\Omega} \lambda(\omega) u(\mu_\omega) \leq \operatorname{cav} u\,(\mu).$$

These inequalities prove that, if $u$ is upper semicontinuous, then so is $\operatorname{cav} u$. As $\operatorname{cav} u$ is a concave function defined on a simplex (namely, $\Delta(\Omega)$), it is automatically also lower semicontinuous (Gale et al. 1968) and is therefore continuous.

In the special case where $\mu^n = \mu$ for all $n \geq 1$, the above inequalities give the equality $\operatorname{cav} u\,(\mu) = \sum_{\omega\in\Omega} \lambda(\omega) u(\mu_\omega)$, which shows that the maximum in (16) is indeed attained. ■

# Appendix B The indispensability of the unidimensionality assumption

The equivalence of conditions *a* and *c* in Theorem 2 does not hold for general utility functions. It is shown below that, in general, (i) *a* does not imply *c*, and (ii) *c* does not imply *a*. Note that finding (i) also implies that, outside the two-state case, the condition in the first part of Theorem 1 is not sufficient. The counterexamples below involve three states, $\Omega = \{\omega_1, \omega_2, \omega_3\}$.

(*a* $\not\Rightarrow$ *c*) Consider the function $\psi: \Delta(\Omega) \longrightarrow \mathbb{R}$ defined by

$$\psi(\mu) = \sqrt{1 - \mu(\omega_1)}\sqrt{\mu(\omega_2)}$$

and the utility function $u$ given by

$$u(\mu) = \begin{cases} \psi(\mu)/2 & \text{if } \mu(\omega_1) > 0 \\ \psi(\mu) & \text{if } \mu(\omega_1) = 0 \end{cases}. \qquad (17)$$

The function $\psi$ is concave and satisfies $\psi \geq u$ everywhere, and therefore $\operatorname{cav} u \leq \psi$. On the other hand, for every $\mu \neq \delta_{\omega_1}$

$$\begin{aligned} \operatorname{cav} u\,(\mu) &= \operatorname{cav} u\left(\mu(\omega_1)\delta_{\omega_1} + \big(1-\mu(\omega_1)\big)\left(\frac{\mu(\omega_2)}{1-\mu(\omega_1)}\delta_{\omega_2} + \frac{\mu(\omega_3)}{1-\mu(\omega_1)}\delta_{\omega_3}\right)\right) \\ &\geq \mu(\omega_1)u\big(\delta_{\omega_1}\big) + \big(1-\mu(\omega_1)\big)u\left(\frac{\mu(\omega_2)}{1-\mu(\omega_1)}\delta_{\omega_2} + \frac{\mu(\omega_3)}{1-\mu(\omega_1)}\delta_{\omega_3}\right) \\ &= \big(1-\mu(\omega_1)\big)\sqrt{\frac{\mu(\omega_2)}{1-\mu(\omega_1)}} = \psi(\mu). \end{aligned}$$

The two inequalities prove that $\operatorname{cav} u = \psi$, and so

$$v(\mu) = \begin{cases} \psi(\mu)/2 & \text{if } \mu(\omega_1) > 0 \\ 0 & \text{if } \mu(\omega_1) = 0 \end{cases}.$$

The function $v$ is concave and therefore condition *a* in Theorem 2 holds. However, condition *c* does not hold, as $u$ is not concave in the entire domain $\Delta(\Omega)$ and is not convex in the interior.

**Remark** A continuous variant of the last example, which is moreover continuously differentiable in the interior of $\Delta(\Omega)$, is

$$u(\mu) = \psi(\mu) - \mu(\omega_1)(1 - \mu(\omega_1)).$$

By the same arguments used above, $\operatorname{cav} u = \psi$, and so $v(\mu)$ is the concave function $\mu(\omega_1)(1 - \mu(\omega_1))$. Yet, in the interior of $\Delta(\Omega)$ the function $u$ is neither concave nor convex, as its Hessian with respect to $\mu(\omega_1)$ and $\mu(\omega_2)$ there has the negative determinant $-0.5\psi(\mu)/\mu(\omega_2)^2$.

(*c* ⇏ *a*) Consider the following variant of the utility function $u$ defined in (17):

$$\breve{u} \coloneqq \max\{u, 1/2\}.$$

The utility function $\breve{u}$ is constant in the interior of $\Delta(\Omega)$ and therefore satisfies condition *c*. However, it follows from the middle part of Theorem 1 (or from Theorem 2 itself) that condition *a* does not hold. This is because in the interior of the face where $\mu(\omega_1) = 0$ we have $\breve{u}(\mu) = \max\left\{\sqrt{\mu(\omega_2)}, 1/2\right\}$, which is not a concave or a convex function.